\RequirePackage{fix-cm}
\documentclass[smallextended]{svjour3}       
\smartqed  
\usepackage{graphicx}
\usepackage{mathptmx}      
\usepackage{latexsym}
\usepackage{amssymb}
\usepackage{dsfont}
 \journalname{Few-Body Systems}
\begin{document}

\title{QCD-constrained dynamical spin effects in the pion holographic light-front wavefunction
\thanks{This research is supported by NSERC Discovery Grants  SAPIN-2017-00031 (R.S) and SAPIN-2017-00033 (M.A).}
}


\author{M. Ahmady \and F. Chishtie \and R. Sandapen        \and
}

\authorrunning{M. Ahmady, F. Chishtie, R. Sandapen} 
\institute{Mohammad Ahmady \at
              Mount Allison University, Sackville, New-Brunswick, E4L 1E6 Canada. \email{mahmady@mta.ca}           
           \and
           Farrukh Chishtie \at
           The University of Western Ontario, London, Ontario, N6A3K7 Canada. \email{fchishti@uwo.ca} 
              \and
           Ruben Sandapen \at
            Acadia University, Wolfville, Nova Scotia, B4P 2R6 Canada. \email{ruben.sandapen@acadiau.ca} \\
             Mount Allison University, Sackville, New-Brunswick, E4L 1E6 Canada. \email{rsandapen@mta.ca}
            }

\date{Received: date / Accepted: date}

\maketitle

\begin{abstract}
Using light-front holography, we  predict simultaneously the pion decay constant and the pion charge radius by taking into account (higher twist) dynamical spin effects whose relative importance is constrained by QCD. 
\end{abstract}

\section{Light-front holography}
\label{intro}

One of the central findings of light-front holographic QCD \cite{deTeramond:2005su,Brodsky:2006uqa,deTeramond:2008ht,Brodsky:2014yha} is the holographic Schr\"odinger equation for mesons : 
 \begin{equation}
			\left(-\frac{\mathrm{d}^2}{\mathrm{d}\zeta^2}-\frac{1-4L^2}{4\zeta^2} + U_{\mathrm{eff}}(\zeta) \right) \phi(\zeta)=M^2 \phi(\zeta) 
	\label{hSE}
	\end{equation}
which is derived within the semiclassical approximation of light-front QCD \cite{Brodsky:2014yha} where quantum loops and quark masses are neglected. The holographic variable 
\begin{equation}
	\mathbf{\zeta}^2 = x(1-x) \mathbf{b}^2_{\perp}
	\end{equation}
maps onto the fifth dimension in anti-de Sitter (AdS) space so that Eq. \ref{hSE} also describes the propagation of weakly-coupled spin-$J$ modes in a modified AdS space. The confining QCD potential is determined by the form of the dilaton field which breaks the conformal invariance of the pure AdS geometry. Specifically, we have
\cite{Brodsky:2014yha}
	\begin{equation}
	U_{\mathrm{eff}}(\zeta)= \frac{1}{2} \varphi^{\prime \prime}(z) + \frac{1}{4} \varphi^{\prime}(z)^2 + \frac{2J-3}{2 z} \varphi^{\prime}(z) \;.
	\label{dilaton-potential}
	\end{equation}	
A remarkable feature in light-front holography is that the form of the confinement potential is uniquely determined \cite{Brodsky:2013ar} to be that of a harmonic oscillator, i.e.  $U_{\mathrm{eff}}=\kappa^4 \zeta^2$. To recover this harmonic potential, the dilaton field has to be quadratic, i.e. $\varphi(z)=\kappa^2 z^2$ so that Eq. \ref{dilaton-potential} then implies that 
	\begin{equation}
		U_{\mathrm{eff}}(\zeta)=\kappa^4 \zeta^2 + 2 \kappa^2 (J-1)	
	\label{hUeff}	
	\end{equation}
	where $J=L+S$. Solving the holographic Schr\"odinger Equation with the confining potential given by Eq. \ref{hUeff} yields the mass spectrum
\begin{equation}
 	M^2= 4\kappa^2 \left(n+L +\frac{S}{2}\right)\;
 	\label{mass-Regge}
 \end{equation}
and the wavefunctions
 \begin{equation}
 	\phi_{nL}(\zeta)= \kappa^{1+L} \sqrt{\frac{2 n !}{(n+L)!}} \zeta^{1/2+L} \exp{\left(\frac{-\kappa^2 \zeta^2}{2}\right)} \nonumber \\ \times ~ L_n^L(\kappa^2 \zeta^2)\;.
 \label{phi-zeta}
 \end{equation}
The immediate striking prediction is that the lowest lying bound state, with quantum numbers $n=L=S=0$, is massless: $M^2=0$.  This state is naturally identified with the pion since the pion mass vanishes in chiral limit $m_f \to 0$.  

The complete meson light-front wavefunction is given by \cite{Brodsky:2014yha}
\begin{equation}
	\Psi(x,\zeta,\varphi)=\frac{\phi(\zeta)}{\sqrt{2\pi \zeta}} X(x) e^{iL\varphi} \;,
	\end{equation}
where $X(x)=\sqrt{x(1-x)}$ \cite{Brodsky:2008pf}. The normalized holographic light-front wavefunction for the pion is then
 \begin{equation}
 	\Psi^{\pi} (x,\zeta^2) = \frac{\kappa}{\sqrt{\pi}} \sqrt{x (1-x)}  \exp{ \left[ -{ \kappa^2 \zeta^2  \over 2} \right] } \;.
\label{pionhwf} 
\end{equation}
The generalization of Eq. \ref{pionhwf} to account for non-vanishing light quark masses is carried out in \cite{Brodsky:2014yha}, yielding
\begin{equation}
\Psi^{\pi} (x,\zeta^2) = \mathcal{N} \sqrt{x (1-x)}  \exp{ \left[ -{ \kappa^2 \zeta^2  \over 2} \right] } \nonumber \\
\times \exp{ \left[ -{m_{f}^2 \over 2 \kappa^2 x(1-x) } \right]}
\label{pion-hwf}
\end{equation}
where $\mathcal{N}$ is a normalization constant which is fixed by requiring that 
 \begin{equation}
 	\int \mathrm{d}^2 \mathbf{b} \mathrm{d} x |\Psi^{\pi}(x,\zeta^2)|^2 = P_{q\bar{q}} 
 	\label{norm}
 \end{equation}
with $P_{q\bar{q}}$ being the probability of finding the pion in the valence Fock sector. Note the light quark masses $m_{f}$ (where $f=u,d$) appearing in the holographic wavefunction are not the current quark masses which appear in the QCD Lagrangian but rather effective quark masses \cite{Brodsky:2014yha}.

Besides quark masses, the other free parameter in light-front holography is the fundamental confinement AdS/QCD scale $\kappa$. Previous work \cite{Forshaw:2012im,Brodsky:2014yha,Brodsky:2016rvj,Deur:2016opc,Ahmady:2016ujw} hints towards a universal value: $\kappa \sim 500$ MeV. Here, we shall use $\kappa=523$ MeV \cite{Deur:2016opc}. In earlier applications of light-front holography with massless quarks, much lower values of $\kappa$ were required to fit the pion data: $\kappa=375$ MeV in Ref. \cite{Brodsky:2007hb} and $\kappa=432$ MeV (with $P_{q\bar{q}}=0.5$) in \cite{Brodsky:2011xx}. In more recent work using constituent quark masses \cite{Vega:2008te,Swarnkar:2015osa,Vega:2009zb,Branz:2010ub}, it turns out that a universal value of $\kappa$ can be used only if the assumption that the pion consists only of the leading valence Fock sector is relaxed:  $P_{q\bar{q}} < 1$. 


\section{Dynamical spin effects in the pion}
Previous work on the pion in light-front holography  was carried out in the leading twist approximation whereby the spin wavefunction decouples from the dynamics. i.e. without dynamical spin effects. To go beyond leading twist, we assume that \cite{Ahmady:2016ufq} 
\begin{equation}
	\Psi(x,\mathbf{k_{\perp}}) \to \Psi(x, \mathbf{k_{\perp}}) S_{\lambda \lambda^{\prime}} (x, \mathbf{k_{\perp}})	
	\label{spin-improved-wf}
	\end{equation}
	with
		\begin{equation}
	S_{\lambda \lambda^{\prime}} (x, \mathbf{k_{\perp}})= \frac{\bar{v}_{\lambda^{\prime}}(x,\mathbf{k_{\perp}})}{\sqrt{1-x}} \left[A \frac{M_{\pi}^2}{P^+} \gamma^+ \gamma^5 + B M_{\pi} \gamma^5 \right] \nonumber \\  \frac{u_{\lambda}(x,\mathbf{k_{\perp}})}{\sqrt{x}} 
	\label{spin-structure} 
	\end{equation}
	where $A$ and $B$ are constants (i.e. momentum-independent)  so that the ratio $\mathcal{R}=A/B$  controls the relative weight of higher twist contributions. In Ref. \cite{Ahmady:2016ufq}, we considered three cases: $[A=1,B=0];[A=0,B=1];[A=B=1]$ without any \textit{\`a priori} theoretical constraint on the ratio $B/A$. However, it is an exact relation in QCD that  \cite{Holl:2004fr,Brodsky:2012ku}  
	\begin{equation}
		f_{\pi} \frac{M_\pi^2}{\hat{m}_{u} + \hat{m}_{d}}=-\langle 0 | \overline{\Psi} \gamma^5 \Psi | \pi \rangle 
	\label{QCD-GMOR}
	\end{equation} 
	where $\hat{m}_{u,d}$ are the renormalized current quark masses. In the chiral limit $\hat{m}_{u,d} \to 0$, Eq. \ref{QCD-GMOR} yields the Gell-Mann-Oakes-Renner (GMOR) relation \cite{GellMann:1968rz}
	\begin{equation}
	M^2_{\pi} = -(\hat{m}_u + \hat{m}_d) \left. \frac{\langle \overline{\Psi}{\Psi} \rangle }{f_{\pi}^2}\right|_{\hat{m}_{u/d} \to 0} 
\label{GMOR}
\end{equation}
 where $\langle \overline{\Psi}{\Psi} \rangle$ is the quark condensate. The GMOR relation is usually interpreted as the vanishing of the pion mass as $M_\pi^2 \propto \hat{m}_{u/d}$ in the chiral limit. On the other hand, if we dare to extrapolate the use of Eq. \ref{GMOR} well beyond the chiral limit, then upon substituting \cite{Ahmady:2016ufq}
	\begin{equation}
	f_{\pi}= 2 \sqrt{\frac{N_c}{\pi}}  \int \mathrm{d} x   \{A((x(1-x) M_{\pi}^2)+ B m_{f} M_{\pi}\} \nonumber \\ \left.  \frac{\Psi^{\pi} (x,\zeta)}{x(1-x)}\right|_{\zeta=0}	 
\label{decayconstant}
\end{equation}
and
\begin{equation}
	\langle 0| \overline{\Psi} \gamma^5 \Psi | \pi \rangle = 2 \sqrt{\frac{N_c}{4\pi}} \int \mathrm{d} x \left[AM_{\pi}^2 m_{f} + \frac{BM_\pi}{x(1-x)}(m_{f}^2 -\nabla_b^2) \right] 
\left. \frac{\Psi^{\pi}(x,\zeta)}{x(1-x)}\right|_{\zeta=0}	
\label{gamma5-ME} 	
\end{equation}
into Eq. \ref{GMOR}, and making use of Eq. \ref{pion-hwf}, we find that
\begin{equation}
	 \frac{B}{A}= -\frac{\int \mathrm{d} x h(x, m_{f},\kappa)}{\int \mathrm{d} x g(x,m_{f},\kappa)}  \equiv \mathcal{R}(m_{f},\kappa)\label{AtoBratio}
\end{equation}	 
	where
	\begin{equation}
			h(x, m_{f}, \kappa) = \frac{1}{\sqrt{x(1-x)}}\left[\frac{m_{f}}{M_\pi} + x(1-x)\left(\frac{M_\pi}{m_{f}}\right) \right] \exp{ \left[ -{m_{f}^2 \over 2 \kappa^2 x(1-x)}  \right] }	
			\end{equation}
			and
	\begin{equation}
		g(x, m_{f},\kappa) = \frac{1}{\sqrt{x(1-x)}}\left[1 + \left(\frac{m_f}{M_\pi}\right)^2\frac{1}{x(1-x)} + 2\left(\frac{\kappa}{M_\pi}\right)^2 \right]\exp{ \left[ -{m_{f}^2 \over 2 \kappa^2 x(1-x)}  \right] }	\;.		
		\end{equation}

In summary, while the chiral limit of the exact QCD relation (Eq. \ref{QCD-GMOR}) yields the GMOR relation, its extrapolation well beyond the chiral limit yields a theoretical constraint on $\mathcal{R}$, i.e. on the relative contribution of dynamical spin effects in the pion holographic wavefunction. 


\section{Predictions for decay constant and charge radius}
 We can now predict the pion decay constant, given by Eq. \ref{decayconstant}, as well as the pion charge radius given by \cite{Ahmady:2016ufq}
\begin{equation}
	\sqrt{\langle r_{\pi}^2 \rangle} = \left[\frac{3}{2} \int \mathrm{d} x \mathrm{d}^2 \mathbf{b} [b (1-x)]^2 |\Psi^{\pi}(x,\mathbf{b})|^2 \right]^{1/2}	
	\label{radius}
	\end{equation}
	where $\Psi^{\pi}(x,\mathbf{b})$ is the two-dimensional Fourier transform of Eq. \ref{spin-improved-wf}. These two observables are interesting since the radius quantifies the departure of the pion from a point-like particle (and thus is sensitive to long-distance physics) while the decay constant, which depends on the wavefunction at zero transverse separation, is sensitive to short-distance physics. A simultaneous description of both observables is therefore challenging (and a stringent test) on any model for the pion light-front wavefunction.

	Our predictions are shown in Table \ref{tab:radius}. As can be seen, the inclusion of QCD-constrained dynamical spin effects suppresses the decay constant while it enhances the radius. The ratio $\mathcal{R}$ which quantifies their importance  does not exceed a few percent but does lead to a very satisfactory description of the data with the larger constituent quark mass. 	
	
\begin{table}
  \centering
  \begin{tabular}{|c|c|c|c|c|c|}
    \hline
Dynamical spin effects &QCD constraint &$m_{f}$ [MeV] &$\mathcal{R}$&$\sqrt{\langle r_{\pi}^2 \rangle}$ [fm]& $f_\pi$ [MeV]\\
        \hline
No & - & $46$ & - &$0.876$ & $162$ \\
        \hline
Yes & Yes &$46$ &$-0.02$ &$1.043$ & $157$\\
   \hline
No & -& $330$ & - &$0.544$ & $161$ \\
        \hline   
Yes & Yes &$330$ & $-0.04$&$0.607$ & $130$\\
   \hline
 Yes & No & $330$ & $1$ &$0.673$ & $138$\\
\hline  
- &-&-& -&$0.672 \pm 0.008$ &$130.4 \pm 0.04 \pm 0.2$\\
    \hline
\end{tabular}
  \caption{Our predictions for the pion charge radius and decay constant using $\kappa=523$ MeV and two different quark masses: $m_{f}=46$ MeV and $m_{f}=330$ MeV, compared to the Particle Data Group averages  \cite{Agashe:2014kda} quoted in the last row. The theoretical prediction with no QCD constraint (and choosing $\mathcal{R}=1$) is from \cite{Ahmady:2016ufq}.}
  \label{tab:radius}
\end{table}
		
\section{Conclusions}
We have taken into account QCD-constrained dynamical spin effects in the holographic pion light-front wavefunction in order to predict simultaneously the pion charge radius and decay constant. We find a remarkable improvement in describing the data when using a constituent quark mass together with the universal AdS/QCD mass scale.   
   
 \section{Acknowledgements}
R.S thanks the organizers of LC2017 for a successful conference in Mumbai and Stan Brodsky for useful discussions.  

\bibliographystyle{spphys}       
\bibliography{sandapen.bib}   

\begin{thebibliography}{10}
\providecommand{\url}[1]{{#1}}
\providecommand{\urlprefix}{URL }
\expandafter\ifx\csname urlstyle\endcsname\relax
  \providecommand{\doi}[1]{DOI \discretionary{}{}{}#1}\else
  \providecommand{\doi}{DOI \discretionary{}{}{}\begingroup
  \urlstyle{rm}\Url}\fi

\bibitem{deTeramond:2005su}
G.F. de~T\'eramond, S.J. Brodsky, Phys. Rev. Lett. \textbf{94}, 201601 (2005).
\newblock \doi{10.1103/PhysRevLett.94.201601}

\bibitem{Brodsky:2006uqa}
S.J. Brodsky, G.F. de~T\'eramond, Phys. Rev. Lett. \textbf{96}, 201601 (2006).
\newblock \doi{10.1103/PhysRevLett.96.201601}

\bibitem{deTeramond:2008ht}
G.F. de~T\'eramond, S.J. Brodsky, Phys. Rev. Lett. \textbf{102}, 081601 (2009).
\newblock \doi{10.1103/PhysRevLett.102.081601}

\bibitem{Brodsky:2014yha}
S.J. Brodsky, G.F. de~T\'eramond, H.G. Dosch, J.~Erlich, Phys. Rept.
  \textbf{584}, 1 (2015).
\newblock \doi{10.1016/j.physrep.2015.05.001}

\bibitem{Brodsky:2013ar}
S.J. Brodsky, G.F. De~T\'eramond, H.G. Dosch, Phys. Lett. \textbf{B729}, 3
  (2014).
\newblock \doi{10.1016/j.physletb.2013.12.044}

\bibitem{Brodsky:2008pf}
S.J. Brodsky, G.F. de~T\'eramond, Phys. Rev. \textbf{D78}, 025032 (2008).
\newblock \doi{10.1103/PhysRevD.78.025032}

\bibitem{Forshaw:2012im}
J.R. Forshaw, R.~Sandapen, Phys. Rev. Lett. \textbf{109}, 081601 (2012).
\newblock \doi{10.1103/PhysRevLett.109.081601}

\bibitem{Brodsky:2016rvj}
S.J. Brodsky, G.F. de~T\'eramond, H.G. Dosch, C.~Lorcé, Int. J. Mod. Phys.
  \textbf{A31}(19), 1630029 (2016).
\newblock \doi{10.1142/S0217751X16300295}

\bibitem{Deur:2016opc}
A.~Deur, S.J. Brodsky, G.F. de~Teramond, J. Phys. \textbf{G44}(10), 105005
  (2017).
\newblock \doi{10.1088/1361-6471/aa888a}

\bibitem{Ahmady:2016ujw}
M.~Ahmady, R.~Sandapen, N.~Sharma, Phys. Rev. \textbf{D94}(7), 074018 (2016).
\newblock \doi{10.1103/PhysRevD.94.074018}

\bibitem{Brodsky:2007hb}
S.J. Brodsky, G.F. de~T\'eramond, Phys. Rev. \textbf{D77}, 056007 (2008).
\newblock \doi{10.1103/PhysRevD.77.056007}

\bibitem{Brodsky:2011xx}
S.J. Brodsky, F.G. Cao, G.F. de~T\'eramond, Phys. Rev. \textbf{D84}, 075012
  (2011).
\newblock \doi{10.1103/PhysRevD.84.075012}

\bibitem{Vega:2008te}
A.~Vega, I.~Schmidt, Phys. Rev. \textbf{D79}, 055003 (2009).
\newblock \doi{10.1103/PhysRevD.79.055003}

\bibitem{Swarnkar:2015osa}
R.~Swarnkar, D.~Chakrabarti, Phys. Rev. \textbf{D92}(7), 074023 (2015).
\newblock \doi{10.1103/PhysRevD.92.074023}

\bibitem{Vega:2009zb}
A.~Vega, I.~Schmidt, T.~Branz, T.~Gutsche, V.E. Lyubovitskij, Phys. Rev.
  \textbf{D80}, 055014 (2009).
\newblock \doi{10.1103/PhysRevD.80.055014}

\bibitem{Branz:2010ub}
T.~Branz, T.~Gutsche, V.E. Lyubovitskij, I.~Schmidt, A.~Vega, Phys. Rev.
  \textbf{D82}, 074022 (2010).
\newblock \doi{10.1103/PhysRevD.82.074022}

\bibitem{Ahmady:2016ufq}
M.~Ahmady, F.~Chishtie, R.~Sandapen, Phys. Rev. \textbf{D95}(7), 074008 (2017).
\newblock \doi{10.1103/PhysRevD.95.074008}

\bibitem{Holl:2004fr}
A.~Holl, A.~Krassnigg, C.D. Roberts, Phys. Rev. \textbf{C70}, 042203 (2004).
\newblock \doi{10.1103/PhysRevC.70.042203}

\bibitem{Brodsky:2012ku}
S.J. Brodsky, C.D. Roberts, R.~Shrock, P.C. Tandy, Phys. Rev. \textbf{C85},
  065202 (2012).
\newblock \doi{10.1103/PhysRevC.85.065202}

\bibitem{GellMann:1968rz}
M.~Gell-Mann, R.J. Oakes, B.~Renner, Phys. Rev. \textbf{175}, 2195 (1968).
\newblock \doi{10.1103/PhysRev.175.2195}

\bibitem{Agashe:2014kda}
K.A. Olive, et~al., Chin. Phys. \textbf{C38}, 090001 (2014).
\newblock \doi{10.1088/1674-1137/38/9/090001}

\end{thebibliography}

\end{document}